\documentclass[RNAAS]{aastex62}

\begin{document}

\title{A Spectroscopic Survey of Superthin Galaxies
\footnote{Published in 2021, {\it Research Notes of the Astronomical Society}, {\bf 5}, 43
\newline \url{https://doi.org/10.3847/2515-5172/abec86}
}}

\correspondingauthor{Stefan Kautsch}
\email{skautsch@nova.edu}

\author{Stefan J. Kautsch}
\altaffiliation{}
\affiliation{Nova Southeastern University, 
Fort Lauderdale, FL 33314}

\author{Dmitry Bizyaev}
\affiliation{Apache Point Observatory and New Mexico State University, Sunspot, NM 88349}
\affiliation{Sternberg Astronomical Institute, Moscow State University, 119992 Moscow, Russia}

\author{Dimitry I. Makarov}
\altaffiliation{}
\affiliation{Special Astrophysical Observatory, 
Russian Academy of Sciences, 
369167 Nizhnij Arkhyz, Russia}

\author{Vladimir P. Reshetnikov}
\affiliation{St. Petersburg State University, 
199034 St. Petersburg, Russia}
\affiliation{Special Astrophysical Observatory, 
Russian Academy of Sciences, 
369167 Nizhnij Arkhyz, Russia}

\author{Alexander V. Mosenkov}
\altaffiliation{}
\affiliation{Central Astronomical Observatory, 
Russian Academy of Sciences, 
196140 St. Petersburg, Russia}

\author{Alexandra V. Antipova}
\altaffiliation{}
\affiliation{Special Astrophysical Observatory, 
Russian Academy of Sciences, 
369167 Nizhnij Arkhyz, Russia}

\begin{abstract}
We present spectroscopic observations of superthin galaxies. 
Superthin galaxies have the thinnest stellar disks among disk 
galaxies. A sample of 138 superthins was observed in visible light 
with the 3.5 m telescope at Apache Point Observatory 
in New Mexico to obtain the rotation curves of the ionized gas in the galaxies. The sample represents the largest survey of superthin galaxies so far and provides a database to investigate the kinematic and dynamic properties of this special type of extragalactic objects. Here we present the rotation curves of our sample objects.
 
\end{abstract} 


\section{Introduction} \label{}

Superthin galaxies are the thinnest disk galaxies. 
Several studies have been conducted over many years because of the 
unusual appearance of this galaxy types, especially from an edge-on perspective, see \citet{goa79}, \citet{kar93}, \citet{kau09}. 
The superthin galaxy types are defined to have a very thin stellar disks, and over the time the definition slightly changed because of differences of defining the extension of galaxies 
(major/minor axes versus scalelength/scaleheight) and the color filter used to image the objects. 
We use a modern definition to select the target galaxies developed by \citet{biz17}. 
In this definition superthins have a scalelength-to-scaleheight ratio of their stellar disk larger than nine ($h/z \geq 9$) in the SDSS r band. 

Superthins are part of the flat, aka simple disk, galaxy class, which is mostly part of Sd Hubble class; and the objects have low surface brightnesses
(\citet{kar93}, \citet{kar99}, \citet{mak01}, \citet{mit05}, \citet{kau06}, \citet{kau09}, \citet{biz14}, \citet{biz17}). 
About 5\% of Sd galaxies are superthins (\citet{kau06}, \citet{biz17}). 
Only very few superthin galaxies have been studied in detail \citet{goa81}, \citet{vdk01}, \citet{mat08}, \citet{obr10}, \citet{biz17}, \citet{ban17}, \citet{kur18}.

Superthin galaxies challenge modern theories of galaxy formation and evolution. Most intriguing is the low efficiency of external or internal dynamical heating processes and the disabled bulge formation (e.g., 
\citet{kor05}, \citet{don06}, \citet{kaz09}, \citet{kho10}, \citet{kan14}, \citet{y19}). Although the buildup of bulges in galaxies has been observed soon after the Big Bang (\citet{lel21}), superthin galaxies do not seem to be affected by this evolution. 
Dark matter halos could play a dominant role in preserving the thin shape of the galaxies, thus, stabilizing the galaxies.
A large sample of spectroscopically observed superthins can provide an insight of possible connections between the properties of the dark halos and the stellar disks. 
Here we report the first results of our long-term spectroscopic survey and present the rotation curves of the objects.

\section{Observations} \label{obs}

We selected the superthins using the same criteria as described in \citet{biz17}. 
The source catalog to select our superthins is the Edge-on Galaxies In SDSS catalog,\footnote{\url{http://users.apo.nmsu.edu/~dmbiz/EGIS/} \newline \url{https://www.sao.ru/edgeon/catalogs.php?cat=EGIS}\newline \url{https://cdsarc.unistra.fr/viz-bin/cat/J/ApJ/787/24}}
\emph{EGIS}, by \citet{biz14}. Our survey contains 138 target objects so far. More superthin galaxies will be added to this survey in the future. The spectra of the targets were collected in visible light with the Dual Imaging Spectrograph on the 3.5 m telescope of the Apache Point Observatory in New
Mexico between December 2014 and August 2019. The spectral resolution is about 5000 by using the high-resolution mode. 
The mean exposure time was 1692 sec (28.2 min) per object. The typical radial velocity accuracy is 12000 m/s (12 km/s). The major
emission lines that are detectable in most targets are: H$\alpha$, H$\beta$, [OIII] 495.9 nm and 500.7 nm, [NII] 654.8 nm and 658.3 nm, as well as [SII] 671.3 nm and 673.2 nm.
The emission line profiles were obtained by fitting a Gaussian distribution. The rotation curves were smoothed with a third degree (cubic) polynomial function.

\begin{center}
\begin{figure}[h!]
\centerline{\includegraphics[scale=1,angle=0]{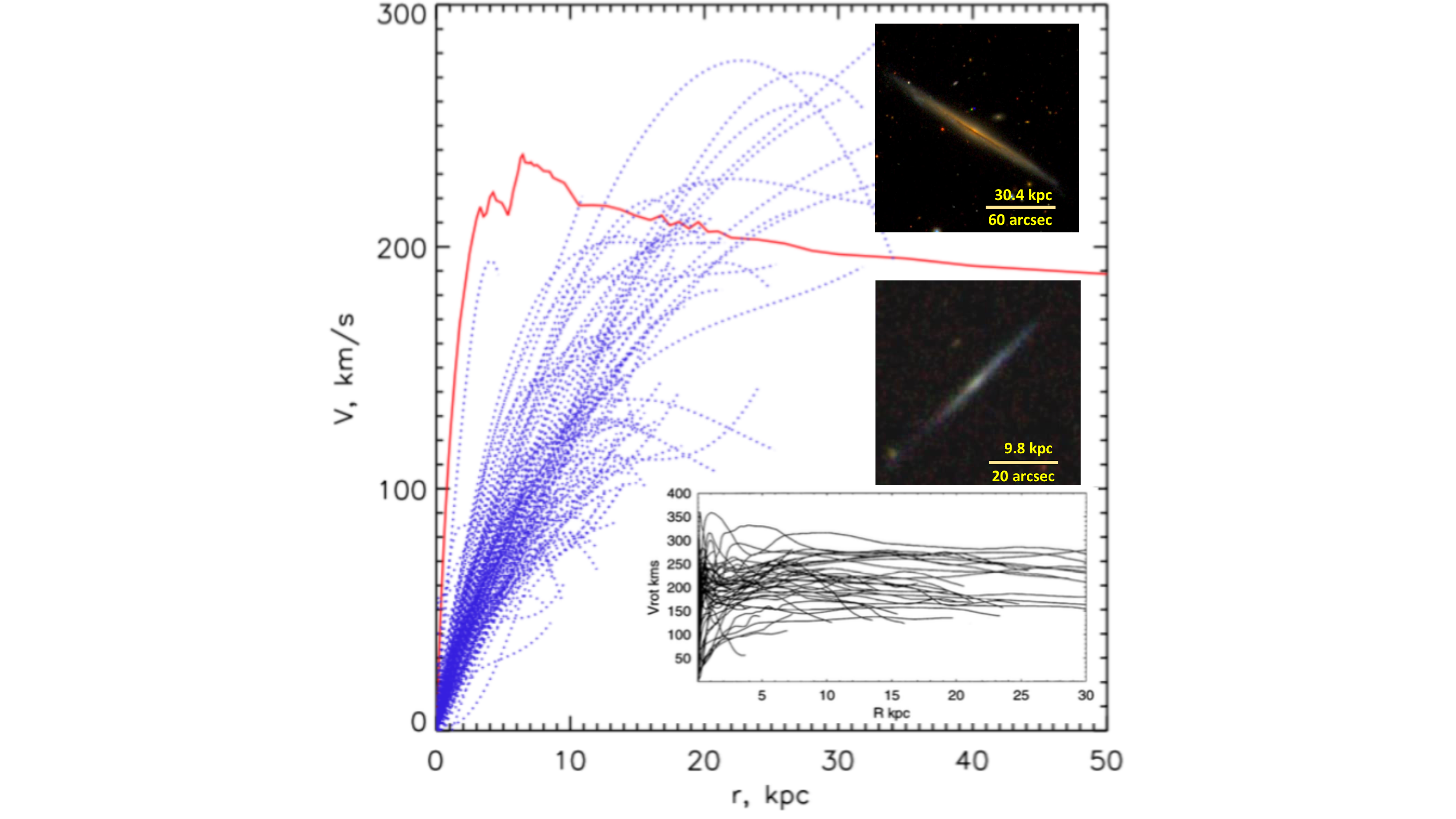}}
\caption{The main figure shows the rotation curves of our 138 galaxies in blue dotted curves. The solid red curve demonstrates the Milky Way
rotation curve. The inlays show representative superthins of a red (upper) and a blue (lower) type, respectively. The horizontal bar in each panel represents the scale in kpc at the distance of the galaxy and the apparent size in arcsec. The bottom inlay shows rotation curves of normal regular disk galaxies from \citet{sof99} for comparison.\label{fig1}}
\end{figure}
\end{center}

Fig. \ref{fig1} shows the rotation curves of the targets. We also plotted our Milky Way's rotation curve from \citet{sof99} for comparison (red solid line), as well as the galactic rotation curves from the \citet{sof99} sample of regular disk galaxies in the bottom inlay of the figure.
The curves of our superthins are steeply rising to the extreme outer parts of the galaxies. This is the opposite of Keplerian curves, and the slopes are much
more comparable to a solid disk with rigid rotation in contrast to the Milky Way and regular disk galaxies. The superthins also exhibit rotation curves with shallow central parts. 
Both phenomena are an indication of a dominant dark matter halo in superthin galaxies. We also observe that the rotation curve maximums are located very far from the galactic centers. This is
a typical feature for low-surface brightness galaxies, which are usually dark-matter dominated objects (e.g., \citet{mcg98}, \citet{zav03}, \citet{biz17}).

In addition, the two other inlays in Fig. \ref{fig1} show two representative superthin galaxies from the EGIS catalog. The upper inlay is an example of a red superthin (EON\_149.150\_20.646). A blue superthin galaxy (EON\_32.766\_6.667) is shown below. The horizontal bar in each panel represents the absolute scale in kpc and the apparent scale in arcseconds. We show these two examples 
because we found in \citet{biz17} that superthins appear in two distinct populations with different properties based on their $(r-i)_0$ color.
The most extreme superthins are blue, have the thinnest disks and lowest surface brightnesses, while the red ones are physically larger (in scalelengths and scaleheights) and are brighter.

\section{Discussion} \label{}

Superthin galaxies are fascinating objects to study and allow us to examine theories of galaxy formation and evolution. 
The most extreme superthins have blue colors and are dynamically underevolved. This means that they are slowly rotating extragalactic
systems, which also exhibit the smallest vertical velocity dispersions, the thinnest stellar disks and lowest surface brightnesses
(see also \citet{biz17}). Redder superthins are rotating faster and are physically larger in all dimensions.

We continue to model the rotation curves including projection effects and the effects of absorption by interstellar matter. Moreover, our model also takes into account the vertical dispersion of the interstellar gas in these
systems. Our preliminary model results tend to support the idea that a dominant, rigid dark halo suppresses internal and external
dynamical heating and therefore prevents disk thickening and bulge formation. However, further calculations are necessary to confirm whether or not a dark halo is solely responsible for squishing the stellar disk in superthin galaxies. 

\section{Acknowledgments} \label{}

This project was partly funded by Nova Southeastern University's President's Faculty Research and Development Grant 335510. The project was also partly supported by grant RSCF grant 19-12-00145.


\begin{thebibliography}{}

\bibitem[Banerjee \& Bapat(2017)]{ban17} Banerjee, A. \& Bapat, D. 2017, \mnras, 466, 3753
\bibitem[Bizyaev et al.(2014)]{biz14} Bizyaev, D., Kautsch, S., Mosenkov, et al.\ 2014, \apj, 787, 24
(EGIS)
\bibitem[Bizyaev et al.(2017)]{biz17} Bizyaev, D., Kautsch, S., Sotnikova, N. Ya., et al.\ 2017, \mnras,
465, 3784
\bibitem[D'Onghia et al.(2005)]{don06} D'Onghia E., Burkert A., Murante G., Khochfar S.\ 2006, \mnras, 372, 1525
\bibitem[Goad \& Roberts(1979)]{goa79} Goad, J. W., \& Roberts, M. S.\ 1979, BAAS, 11, 668
\bibitem[Goad \& Roberts(1981)]{goa81} Goad, J. W., \& Roberts, M. S.\ 1981, \apj, 250, 79
\bibitem[Jadhav Y \& Banerjee(2019)]{y19} Jadhav Y, V., \& Banerjee, A.\ 2019, \mnras, 488, 547
\bibitem[Kanak(2014)]{kan14} Kanak, S. 2014, arXiv, 1403.1711S
\bibitem[Karachentsev et al.(1993)]{kar93} Karachentsev, I. D., Karachentseva, V. E., Parnovskij, S. L.\ 1993,
Astronomische Nachrichten, 314, 97
\bibitem[Karachentsev et al.(1999)]{kar99} Karachentsev, I. D., Karachentseva, V. E., Kudrya, Yu. N., et al.\ 1999, Bull. Spec. Astr. Obs., 47, 5
\bibitem[Kautsch et al.(2006)]{kau06} Kautsch, S. J., Grebel, E. K., Barazza, F. D., \& Gallagher, J. S.,
III\ 2006, A\&A, 445, 765
\bibitem[Kautsch(2009)]{kau09} Kautsch, S. J.\ 2009, \pasp, 121, 1297
\bibitem[Kazantzidis et al.(2009)]{kaz09} Kazantzidis, S., Zentner, A. R., Kravtsov, A. V., Bullock, J. S. et
al.\ 2009, \apj, 700, 1896
\bibitem[Khoperskov et al.(2010)]{kho10} Khoperskov, A., Bizyaev, D., Tiurina, N., \& Butenko, M.\ 2010,
Astronomische Nachrichten, 331, 731
\bibitem[Kormendy et al.(2005)]{kor05} Kormendy, J. \& Fisher, D. B.\ 2005, RMXAA (Conf. Ser.), 23,
101
\bibitem[Kurapati et al.(2018)]{kur18} Kurapati, S. et al.\ 2018, \mnras, 479, 5686 
\bibitem[Lelli et al.(2021)]{lel21} Lelli, F. et al.\ 2021, Science, 371, 713
\bibitem[Makarov et al.(2001)]{mak01} Makarov, D. I., Burenkov, A. N., \& Tyurina, N. V.\ 2001, Astron.
Letters, 27, 213
\bibitem[Matthews \& Uson(2008)]{mat08} Matthews, L. D. \& Uson, J. M. 2008, \aj, 135, 291
\bibitem[McGaugh \& de Blok(1998)]{mcg98} McGaugh, S. S., \& de Blok, W. J. G. 1998, \apj, 499, 41
\bibitem[Mitronova et al.(2005)]{mit05} Mitronova, S. N., Huchtmeier, W. K., Karachentsev, I. D.,
Karachentseva, V. E., et al.\ 2005, Astron. Letters, 31, 501
\bibitem[O'Brien et al.(2010)]{obr10} O'Brien, J. C., et al. 2010, A\&A, 515, 62
\bibitem[Sofue et al.(1999)]{sof99} Sofue, Y., Tutui, Y., Honma, M. et al.\ 1999, \apj, 523, 136
\bibitem[van der Kruit et al.(2001)]{vdk01} van der Kruit, P. C., et al. 2001, A\&A, 379, 374
\bibitem[Zavala et al.(2003)]{zav03} Zavala, J., Avila-Reese, V., Hernandez-Toledo, H., Firmani, C. 2003, A\&A 412, 633 

\bibitem[()]{}
\end{thebibliography}
\end{document}